\documentclass[10pt]{article}
\begin{document}
\pagestyle{plain}
\setcounter{page}{1}
\begin{center}
{\large\bf Noncommutative Quantum Gravity}
\vskip 0.3 true in
{\large J. W. Moffat}
\vskip 0.3 true in
{\it Department of Physics, University of Toronto,
Toronto, Ontario M5S 1A7, Canada}
\end{center}
\begin{abstract}%
The possible role of gravity in a noncommutative geometry is investigated.
Due to the Moyal *-product of fields in noncommutative geometry, it is
necessary to complexify the metric tensor of gravity. We first consider
the possibility of a complex Hermitian, nonsymmetric $g_{\mu\nu}$ and
discuss the problems associated with such a theory. We then introduce a
complex symmetric (non-Hermitian) metric, with the associated
complex connection and curvature, as the basis of a noncommutative
spacetime geometry. The spacetime coordinates are in general complex and
the group of local gauge transformations is associated with the complex
group of Lorentz transformations $CSO(3,1)$. A real action is chosen to
obtain a consistent set of field equations. A Weyl quantization of the
metric associated with the algebra of noncommuting coordinates is employed.
\end{abstract}



\section{Introduction}

The concept of a quantized spacetime was proposed by Snyder~\cite{Snyder}, and has
received much attention over the past few
years [2-6]. There has been renewed interest
recently in noncommutative field theory, since it makes its appearance
in string theory, e.g. noncommutative gauge theories describe the low
energy excitations of open strings on D-branes in a background two-form B
field [5,7-12]. Noncommutative Minkowski space is defined in
terms of spacetime coordinates $x^\mu,\mu=0,...3$, which satisfy
the following commutation relations
\begin{equation}
[x^\mu,x^\nu]=i\theta^{\mu\nu},
\end{equation}
where $\theta^{\mu\nu}$ is an antisymmetric tensor. In what
follows, we can generally extend the results to higher dimensions
$\mu=0,...d.$

An associative noncommuting algebra, $M^4_{\theta}$, is constructed with
elements given by ordinary continuous functions on $M^4$ and with a
deformed product of functions given by the Moyal bracket or $\star$-product
of functions~\cite{Moyal}
\begin{equation}
f(x)\star g(x)
=\exp\biggl[\frac{i}{2}\theta^{\mu\nu}\frac{\partial}{\partial\alpha^\mu}
\frac{\partial}{\partial\beta^\nu}\biggr]f(x+\alpha)g(x+\beta)\vert_{\alpha=\beta=0}
$$
$$
=f(x)g(x)+\frac{i}{2}\theta^{\mu\nu}\partial_\mu f(x)\partial_\nu
g(x)+{\cal O}(\theta^2).
\end{equation}
Conventional field theories are generalized to noncommutative spacetime
by replacing the usual product of fields by the Moyal bracket or $\star$-product.

Since the $\star$-product of fields involves an infinite number
of derivatives, the resulting field theories are nonlocal. The lack of commutativity of
the spacetime coordinates gives rise to a spacetime uncertainty
relation 
\begin{equation}
\Delta x^\mu\Delta x^\nu \geq \frac{1}{2}\vert\theta^{\mu\nu}\vert.
\end{equation}
It appears that in a perturbative context, the noncommutative theories have
a unitary S-matrix for space-space noncommutativity, 
$\theta^{ij}\not= 0,\theta^{0i}=0$, while the S-matrix is not unitary for
space-time noncommutativity, $\theta^{0i}\not= 
0,\theta^{ij}=0$~\cite{Gomis}. Moreover, there are indications that 
conventional renormalizable field theories remain renormalizable, when 
generalized to noncommutative spacetimes~\cite{Varilly,Martin}.

The product of two operators ${\hat
f}$ and ${\hat g}$ can be defined and can be shown to lead to the Moyal
$\star$-product~\cite{Wess}
\begin{equation}
f(x)\star g(x)=\frac{1}{(2\pi)^4}\int
d^4kd^4p\exp[i(k_\mu+p_\mu)x^\mu-\frac{i}{2}k_\mu\theta^{\mu\nu}p_\nu]{\tilde
f}(k){\tilde g}(p)
$$
$$
=\exp\biggl[\frac{i}{2}\frac{\partial}{\partial
x^\mu}\theta^{\mu\nu}\frac{\partial}{\partial
y^\nu}\biggr]f(x)g(y)\mid_{y\rightarrow x},
\end{equation}
where ${\tilde f}(k)$ is the Fourier transform
\begin{equation}
{\tilde f}(k)=\frac{1}{(2\pi)^2}\int d^4x\exp(-ik_\sigma x^\sigma)f(x).
\end{equation}

The noncommutative Yang-Mills action is defined by
\begin{equation}
S_{YM}=-\frac{1}{4}\int d^4xF^a_{\mu\nu}\star F^{a\mu\nu},
\end{equation}
where
\begin{equation}
\label{F-field}
F^a_{\mu\nu}=\partial_\mu A^a_\nu-\partial_\nu A^a_\mu+i\epsilon^{abc}A_{b\mu}\star
A_{c\nu}.
\end{equation}
The action is invariant under the gauge transformations
\begin{equation}
A^a_\mu\rightarrow [U\star A_\mu\star U^{-1}-\partial_\mu U\star U^{-1}]^a,
\end{equation}
where
\begin{equation}
U\star U^{-1}=U^{-1}\star U=1.
\end{equation}

The definition (\ref{F-field}) of $F^a_{\mu\nu}$ requires the gauge fields
to be complex and these fields should be invariant under the
transformations of a general group of noncommutative gauge transformations
$NCU(3,1)$. When we contemplate a noncommutative extension of spacetime,
we do not appear to be able to proceed with the quantization of fields as
in commutative quantum field theory, in which the classical fields are real
for neutral charged fields and complex for charged fields. Moreover, the
classical field variables are treated in the quantization procedure as
operators in a Hilbert space, and they satisfy commutation or
anticommutation relations. Within this standard scenario, spacetime plays a
passive role as far as quantization is concerned.

Let us now consider the role of gravity in a noncommutative
spacetime. We can employ the Weyl quantization
procedure~\cite{Wess,Weyl} to associate the metric tensor
operator ${\hat g}_{\mu\nu}$ with the classical metric
$g_{\mu\nu}$. This prescription can be used to associate an
element of the noncommutative algebraic structure $M_x$, which
defines a noncommutative space in terms of the coordinate
operators ${\hat x}$. Using the Fourier transform of the metric
\begin{equation} {\tilde g}_{\mu\nu}(k)=\frac{1}{(2\pi)^2}\int
d^4x \exp(-ik_\sigma x^\sigma)g_{\mu\nu}(x), \end{equation}
we can define the metric operator
\begin{equation}
{\hat g}_{\mu\nu}(\hat x)=\frac{1}{(2\pi)^2}\int d^4k\exp(ik_\sigma{\hat
x}^\sigma){\tilde g}_{\mu\nu}(k).
\end{equation}
Thus, the operators ${\hat g}_{\mu\nu}$ and ${\hat x}$ replace the variables
$g_{\mu\nu}$ and $x$. If the ${\hat x}$ have complex
symmetry properties, then the ${\hat g}_{\mu\nu}$ will inherit these
properties for a classical metric $g_{\mu\nu}$.

When we turn to the geometry of spacetime in the
presence of matter, the situation with regards to noncommutativity of
spacetime coordinates is much more problematic. In this case, it seems
unlikely that we can retain our conventional notions of a real
pseudo-Riemannian space within a real manifold. If we define the metric of
spacetime in terms of vierbeins $v^a_\mu$ using the Moyal $\star$-product
\begin{equation}
g_{\mu\nu}=v^a_\mu\star v^b_\nu\eta_{ab},
\end{equation}
where $a,b= 0,...3$ denote the flat, fiber bundle tangent space
(anholonomic) coordinates, and $\eta_{ab}$ denotes the Minkowski metric:
$\eta_{ab}={\rm diag}(1, -1, -1, -1)$, then the metric of spacetime is
forced to be complex.

We shall investigate two possibilities for complex gravity.
First we consider the possibility that the fundamental tensor $g_{\mu\nu}$
is complex Hermitian satisfying $g_{\mu\nu}^\dagger=g_{\mu\nu}$, where
$\dagger$ denotes the Hermitian conjugate. This was recently proposed by
Chamseddine~\cite{Chamseddine,Majid}, as a possible way to complexify the
gravitational metric. The second possibility we shall consider is a
complex pseudo-Riemannian geometry, based on a complex symmetric
(non-Hermitean) tensor $g_{\mu\nu}$~\cite{Moffat,Moffat2,Moffat3}.

\section{Nonsymmetric Gravity}

The nonsymmetric field extension of Einstein gravity has a long history.
It was originally proposed by Einstein, as a unified field theory of gravity and
electromagnetism~\cite{Einstein,Straus}. But it was soon realized that the
antisymmetric part $g_{[\mu\nu]}$ in the decomposition
\begin{equation}
\label{nonsymmetric}
g_{\mu\nu}=g_{(\mu\nu)}+g_{[\mu\nu]}
\end{equation}
could not describe physically the electromagnetic field. It was then
suggested that the nonsymmetric field structure describes a
generalization of Einstein gravity, known in the literature as the nonsymmetric
gravitational theory
(NGT) [27-32].

NGT faces some difficulties which have their origin in the lack of a clear-cut gauge
symmetry in the antisymmetric sector of the theory. Even in the linear
approximation, the antisymmetric field equations are not invariant under
the transformation
\begin{equation}
g^\prime_{[\mu\nu]}=
g_{[\mu\nu]}+\partial_\nu\lambda_\mu-\partial_\mu\lambda_\nu,
\end{equation}
where $\lambda_\mu$ is an arbitrary vector field. This lack of
overall gauge symmetry in the antisymmetric sector of the theory
gives rise to two basic problems~\cite{Damour,Clayton,Clayton2}.
In the weak antisymmetric field approximation, an expansion about
a classical general relativity (GR) background, reveals that
there are ghost poles, tachyons and higher-order poles associated
with the asymptotic boundary conditions~\cite{Damour}. However,
this problem can be resolved by a careful choice of the NGT
action~\cite{Moffat8,Moffat9}: \begin{equation}
\label{NGTaction}
S_{NGT}=\int
d^4x\sqrt{-g}[g^{\mu\nu}R_{\mu\nu}(W)-2\lambda-\frac{1}{4}\mu^2g^{\mu\nu}g_{[\nu\mu]}
-\frac{1}{6}g^{\mu\nu}W_\mu W_\nu].
\end{equation}
Here, we choose units so that $G=c=1$, $g={\rm Det}(g_{\mu\nu})$, $\lambda$
is the cosmological constant, $\mu$ is a mass associated with the
skew field $g_{[\mu\nu]}$ and $R_{\mu\nu}(W)$ is the contracted curvature
tensor:
\begin{equation}
R_{\mu\nu}(W)=\partial_\beta
W^\beta_{\mu\nu}-\frac{1}{2}(\partial_\nu W^\beta_{\mu\beta}+\partial_\mu
W^\beta_{\nu\beta})-W^\beta_{\alpha\nu}W^\alpha_{\mu\beta}
+W^\beta_{\alpha\beta}W^\alpha_{\mu\nu},
\end{equation}
defined in terms of the unconstrained nonsymmetric connection
\begin{equation}
\label{Wconnection}
W^\lambda_{\mu\nu}=\Gamma^\lambda_{\mu\nu}-\frac{2}{3}\delta^\lambda_\mu
W_\nu, \end{equation}
where
\begin{equation}
W_\mu=\frac{1}{2}(W^\lambda_{\mu\lambda}-W^\lambda_{\lambda\mu}).
\end{equation}
Eq.(\ref{Wconnection}) leads to the result
\begin{equation}
\Gamma_\mu=\Gamma^\lambda_{[\mu\lambda]}=0.
\end{equation}
A nonsymmetric matter source is added to the action
\begin{equation}
S_M=-8\pi\int d^4xg^{\mu\nu}\sqrt{-g}T_{\mu\nu},
\end{equation}
where $T_{\mu\nu}$ is a nonsymmetric source tensor.

This action will lead to a physically consistent Lagrangian and field
equations for the antisymmetric field in the linear, weak field
approximation, with field equations of the form of a massive Proca-type
theory which is free of tachyons, ghost poles and higher-order
poles~\cite{Moffat8,Moffat9}.

However, the problems do not end here with this form of NGT. A Hamiltonian
constraint analysis performed on NGT by Clayton, showed that when the NGT field
equations are expanded about a classical GR background, the resulting theory is
unstable~\cite{Clayton,Clayton2}. This problem appears to be a generic
feature of any fully geometrical NGT-type of theory, including the vierbein
derivation of NGT based on the Hermitian fundamental
tensor~\cite{Moffat6,Moffat7,Chamseddine}
\begin{equation}
g_{\mu\nu}=e^a_\mu {\tilde e}^{b}_\nu\eta_{ab},
\end{equation}
where $e^a_\mu$ is a complex vierbein and ${\tilde e}^a_\mu$ denotes the 
complex conjugate of $e^a_\mu$. Basically, this result means that we cannot
consider GR as a sensible limit of NGT for weak antisymmetric fields.

Another serious problem with the complex version of NGT, in which
$g_{[\mu\nu]}=if_{[\mu\nu]}$ and $f_{[\mu\nu]}$ is a real antisymmetric
tensor, is that the linear, weak field approximation to the NGT field
equations produces generic, negative energy ghost poles. This prompted a
proposal that the nonsymmetric vierbeins be described by hyperbolic complex
variables~\cite{Moffat6,Moffat7}. For a sesquilinear, hyperbolic complex
$g_{\mu\nu}$, there exists a local $GL(4,R)$ gauge symmetry, which
corresponds to $g_{\mu\nu}$ preserving rotations of generalized linear
frames in the tangent bundle. This symmetry should not be confused with the
linear (global) subgroup $GL(4)$ of the diffeomorphism group of the
manifold ${\cal M}^4$ under which NGT is also invariant. The hyperbolic
complex vierbeins are defined by
\begin{equation}
e^a_\mu={\rm Re}(e^a_\mu)+\omega{\rm Im}(e^a_\mu),
\end{equation}
while the $g_{\mu\nu}$ is given by
\begin{equation}
g_{\mu\nu}=e^a_\mu{\tilde e}^b_\nu\eta_{ab}=e^a_\mu{\tilde e}_{\nu a},
\end{equation}
where $\omega^2=+1$ is the pure imaginary element of the hyperbolic complex
Clifford algebra $\Omega$~\cite{Clifford}. The $g_{\mu\nu}$ and the
connexion $\Gamma^\lambda_{\mu\nu}$ are hyperbolic complex Hermitian in
$\mu$ and $\nu$, while the spin connection $(\Omega_\sigma)_{ab}$ is
hyperbolic complex skew-Hermitian in $a$ and $b$. The hyperbolic complex
unitary group $U(3,1,\Omega)$ is isomorphic to $GL(4,R)$. The spin
connection $(\Omega_\sigma)^a_b$ is invariant under the $GL(4)$
transformations provided \begin{equation}
(\Omega_\sigma)^a_b\rightarrow[U_G\Omega_\sigma U_G^{-1}-(\partial_\sigma
U_G)U_G^{-1}]^a_b,
\end{equation}
where $U_G$ is an element of the unitary group $U(3,1,\Omega)$. The
curvature tensor $(R_{\mu\nu})^a_b$ is invariant under the $GL(4)$
transformations when
\begin{equation}
(R_{\mu\nu})^a_b\rightarrow U^a_{Gc}(R_{\mu\nu})^c_d(U_G^{-1})^d_b.
\end{equation}

The field equations can be found from the action~\cite{Moffat6,Moffat7}
\begin{equation}
S_{\rm grav}=\int d^4xeR(e),
\end{equation}
where $e=\sqrt{-g}$ and $R=e^{\mu a}{\tilde e}^{\nu b}(R_{\mu\nu})_{ab}$.
Although the particle spectrum is now free of negative energy ghost states
in the weak field approximation, the theory still suffers from the
existence of dipole ghost states due to the asymptotic boundary conditions
for $g_{[\mu\nu]}$, unless we use the action (\ref{NGTaction}),
rewritten in the language of vierbeins. However, a Hamiltonian
constraint analysis for this theory will still reveal serious
instability problems~\cite{Clayton,Clayton2}.

\section{Complex Symmetric Riemannian Geometry}

We shall now consider choosing a complex manifold of coordinates ${\cal
M}_C^4$ and a complex symmetric metric defined
by~\cite{Moffat,Moffat2,Moffat3}
\begin{equation}
g_{\mu\nu}=s_{\mu\nu}+a_{\mu\nu},
\end{equation}
where $a_{\mu\nu}=ib_{\mu\nu}$ and $b_{\mu\nu}$ is a real symmetric
tensor. This corresponds to having two copies of a real metric in an
eight-dimensional space. The real diffeomorphism symmetry of standard
Riemannian geometry is extended to a complex diffeomorphism symmetry under
the group of complex coordinates transformations with $z^\mu=x^\mu+iy^\mu$.
The metric can be expressed in terms of a complex vierbein $E^a_\mu={\rm
Re}(E^a_\mu)+i{\rm Im}(E^a_\mu)$ as
\begin{equation}
\label{complexvierbein}
g_{\mu\nu}=E^a_\mu E^b_\nu\eta_{ab}.
\end{equation}

The real contravariant tensor $s^{\mu\nu}$ is associated with $s_{\mu\nu}$
by the relation
\begin{equation}
s^{\mu\nu}s_{\mu\sigma}=\delta^\nu_\sigma
\end{equation}
and also
\begin{equation}
g^{\mu\nu}g_{\mu\sigma}=\delta^\nu_\sigma.
\end{equation}
With the complex spacetime is also associated a complex symmetric
connection
\begin{equation}
\Gamma^\lambda_{\mu\nu}=\Delta^\lambda_{\mu\nu}+\Omega^\lambda_{\mu\nu},
\end{equation}
where $\Omega^\lambda_{\mu\nu}$ is purely imaginary.

We shall determine the $\Gamma^\lambda_{\mu\nu}$ according to the
$g_{\mu\nu}$ by the equations
\begin{equation}
\label{gequation}
g_{\mu\nu;\lambda}=\partial_\lambda
g_{\mu\nu}-g_{\rho\nu}\Gamma^\rho_{\mu\lambda}-g_{\mu\rho}\Gamma^\rho_{\nu\lambda}=0.
\end{equation}
By commuting the two covariant differentiations of an arbitrary complex
vector $A_\mu$, we obtain the generalized curvature tensor
\begin{equation}
R^\lambda_{\mu\nu\sigma}=-\partial_\sigma\Gamma^\lambda_{\mu\nu}
+\partial_\nu\Gamma^\lambda_{\mu\sigma}+\Gamma^\lambda_{\rho\nu}\Gamma^\rho_{\mu\sigma}
-\Gamma^\lambda_{\rho\sigma}\Gamma^\rho_{\mu\nu}
\end{equation}
and a contracted curvature tensor $R_{\mu\nu}=R^\sigma_{\mu\nu\sigma}$:
\begin{equation}
R_{\mu\nu}=A_{\mu\nu}+B_{\mu\nu},
\end{equation}
where $B_{\mu\nu}$ is a purely imaginary tensor.
From the curvature tensor, we can derive the four complex (eight real)
Bianchi identities \begin{equation}
\label{Bianchi}
(R^{\mu\nu}-\frac{1}{2}g^{\mu\nu}R)_{;\nu}=0.
\end{equation}

We must choose a real action to guarantee a consistent set of field
equations. There is, of course, a degree of arbitrariness in choosing this
action due to the complex manifold of coordinate transformations and the
complex Riemannian geometry. We shall choose~\cite{Moffat2}
\begin{equation}
\label{action}
S_{\rm grav}=\frac{1}{2}\int d^4x[{\bf g}^{\mu\nu}R_{\mu\nu}+{\rm
compl. conj.}]=\int d^4x[{\bf s}^{\mu\nu}A_{\mu\nu}
+{\bf a}^{\mu\nu}B_{\mu\nu}],
\end{equation}
where ${\bf g}^{\mu\nu}=\sqrt{-g}g^{\mu\nu}={\bf s}^{\mu\nu}+{\bf
a}^{\mu\nu}$. The variation with respect to ${\bf s}^{\mu\nu}$ and ${\bf
a}^{\mu\nu}$ yields the twenty field equations of empty space
\begin{equation}
A_{\mu\nu}=0,\quad B_{\mu\nu}=0,
\end{equation}
or, equivalently, the ten complex field equations
\begin{equation}
\label{fieldequations}
R_{\mu\nu}=0.
\end{equation}
We can add a real matter action to (\ref{action}):
\begin{equation}
S_{\rm matter}=-4\pi\int d^4x[{\bf g}^{\mu\nu}T_{\mu\nu}+{\rm compl.
conj.}],
\end{equation}
where $T_{\mu\nu}=S_{\mu\nu}+C_{\mu\nu}$ is a complex symmetric
source tensor, and $S_{\mu\nu}$ and $C_{\mu\nu}$ are real and pure
imaginary tensors, respectively.

We shall assume that the line element determining the physical
gravitational field is of the real form
\begin{equation}
\label{realmetric}
ds^2=s_{\mu\nu}dx^\mu dx^\nu.
\end{equation}
Let us consider the static spherically symmetric line element
\begin{equation}
ds^2=\alpha dt^2-\eta dr^2-r^2(d\theta^2+\sin^2\theta d\phi^2),
\end{equation}
where the real $\alpha$ and $\eta$ are functions of real $r$
only. Our complex fundamental tensor $g_{\mu\nu}$ is determined
by \begin{equation}
g_{11}(r)\equiv -\mu(r)=-[\eta(r)+i\zeta(r)],
$$
$$
g_{22}(r)\equiv s_{22}(r)=-r^2,
$$
$$
g_{33}(r)\equiv s_{33}(r)=-r^2\sin^2\theta,
$$
$$
g_{00}(r)\equiv \gamma(r)=\alpha(r)+i\beta(r).
\end{equation}
Solving the $\Gamma^\lambda_{\mu\nu}$ from Eq.(\ref{gequation}) and substituting
into the field equation (\ref{fieldequations}), we get the solutions
\begin{equation}
\alpha=1-\frac{2m}{r},\quad \beta=\frac{2\epsilon}{r},
\end{equation}
where $2m$ and $2\epsilon$ are constants of integration, and we have imposed the
boundary conditions
\begin{equation}
\alpha\rightarrow 1,\quad \eta\rightarrow 1,\quad \beta\rightarrow 0,\quad
\zeta\rightarrow 0,
\end{equation}
as $r\rightarrow\infty$. Solving for $\eta(r)$ and $\zeta(r)$ from the
solution $\mu(r)=1/\gamma(r)$ we get~\cite{Moffat3}
\begin{equation}
ds^2= (1-\frac{2m}{r})dt^2
-\frac{1-\frac{2m}{r}}{(1-\frac{2m}{r})^2+\frac{4\epsilon^2}{r^2}}dr^2
-r^2(d\theta^2+\sin^2\theta d\phi^2),
\end{equation}
\begin{equation}
\zeta=-\frac{2\epsilon/r}{(1-\frac{2m}{r})^2+\frac{4\epsilon^2}{r^2}}.
\end{equation}
When $\epsilon\rightarrow 0$, we regain the Schwarzschild solution of Einstein
gravity.

The weak field approximation obtained from the expansion about Minkowski
spacetime
\begin{equation}
g_{\mu\nu}=\eta_{\mu\nu}+h_{\mu\nu},
\end{equation}
where $h_{\mu\nu}=p_{\mu\nu}+ik_{\mu\nu}$, leads to an action
which is invariant under the local, linear gauge transformation
\begin{equation}
h^\prime_{\mu\nu}=h_{\mu\nu}+\partial_\mu\theta_\nu+\partial_\nu\theta_\mu,
\end{equation}
where $\theta_\mu$ is an arbitrary complex vector. Thus, two spin 2 massless
gravitons describe the complex bimetric gravity field. The physical spin 2 graviton
is described by a real mixture of the basic spin two particles associated with
$p_{\mu\nu}$ and $k_{\mu\nu}$. Moreover, the physical null cone with
$ds^2=0$ will be a real mixture of the two null cones associated with
$s_{\mu\nu}$ and $a_{\mu\nu}$.

The linearized action $S_{\rm grav}$, obtained from (\ref{action})
in the weak field approximation, contains negative energy ghost states
coming from the purely imaginary part of the metric $k_{\mu\nu}$. To avoid
this unphysical aspect of the theory, we can base the geometry on a
hyperbolic complex metric
\begin{equation}
g_{\mu\nu}=s_{\mu\nu}+\omega a_{\mu\nu},
\end{equation}
where $\omega$ is the pure imaginary element of a hyperbolic complex
Clifford algebra $\Omega$~\cite{Clifford}. We have $\omega^2=+1$ and for
$z=x+\omega y$, we obtain $z{\tilde z}=x^2-y^2$ where ${\tilde z}=x-\omega 
y$, so that there exist lines of zeros, $z{\tilde z}=0$, in the hyperbolic
complex space and it follows that $\Omega$ forms a ring of numbers and not
a field as for the usual system of complex numbers with the pure imaginary
element $i=\sqrt{-1}$. The linearized action $S_{\rm grav}$ should now have
positive energy and with the invariance under the hyperbolic complex group
of transformations $CSO(3,1,\Omega)$ the gravitons should be free of ghost
states.

\section{Noncommutative Complex Symmetric Gravity}

We must now generalize the complex symmetric gravity (CSG) to
noncommutative coordinates by replacing the usual
products by Moyal-products. We shall use the complex vierbein
formalism based on the metric (\ref{complexvierbein}) and a complex spin
connection $(\omega_\mu)_{ab}$, since this formalism is closer to the
standard gauge field formalism of field theory.

In order to accommodate a complex symmetric metric, we shall treat the
coordinates as ``fermionic'' degrees of freedom and use the anticommutation
relation
\begin{equation}
\label{anticommutator}
\{{\hat x}^\mu,{\hat x}^\nu\}=2x^\mu x^\nu+i\tau^{\mu\nu},
\end{equation}
where $\tau^{\mu\nu}=\tau^{\nu\mu}$ is a symmetric tensor.
The Moyal product of two operators is now given by the
$\diamond$-product
\begin{equation}
f(x)\diamond g(x)
=\exp\biggl[\frac{i}{2}\tau^{\mu\nu}\frac{\partial}{\partial\alpha^\mu}
\frac{\partial}{\partial\beta^\nu}\biggr]f(x+\alpha)g(x+\beta)\vert_{\alpha=\beta=0}
$$ $$
=f(x)g(x)+\frac{i}{2}\tau^{\mu\nu}\partial_\mu f(x)\partial_\nu g(x)+{\cal
O}(\tau^2).
\end{equation}
If we form $x^\mu\diamond x^\nu$ and add it to $x^\nu\diamond
x^\mu$, then we obtain the anticommutator expression
(\ref{anticommutator}).

The complex symmetric metric is defined by
\begin{equation}
g_{\mu\nu}=E^a_\mu\diamond E^b_\nu\eta_{ab}.
\end{equation}
The spin connection is subject to the gauge
transformation
\begin{equation}
(\omega_\sigma)^a_b\rightarrow[U_C\diamond\omega_\sigma\diamond
U_C^{-1}-(\partial_\sigma U_C)\diamond U_C^{-1}]^a_b,
\end{equation}
where $U_C$ is an element of a complex noncommutative group of
orthogonal transformations $NCSO(3,1)$.

The curvature tensor is now given by
\begin{equation}
(R_{\mu\nu})^a_b=\partial_\mu(\omega_\nu)^a_b-\partial_\nu(\omega_\mu)^a_b
+(\omega_\mu)^a_c\diamond(\omega_\nu)^c_b-(\omega_\nu)^a_c\diamond
(\omega_\mu)^c_b,
\end{equation}
which transforms as
\begin{equation}
(R_{\mu\nu})^a_b\rightarrow
U^a_{Cc}\diamond(R_{\mu\nu})^c_d\diamond(U_C^{-1})^d_b.
\end{equation}

The real action is
\begin{equation}
S_{\rm grav}=\frac{1}{2}\int d^4x[E\diamond E^\mu_a\diamond
(R_{\mu\nu})^a_b\diamond E^{\nu b} + {\rm compl. conj.}],
\end{equation}
where $E=\sqrt{-g}$.  The action $S_{\rm grav}$ is locally invariant under the
transformations of the complex noncommutative, fiber bundle tangent space
group $NCSO(3,1)$, i.e. the group of complex noncommutative homogeneous
Lorentz transformations. It has been shown by Bonora et al.,~\cite{Bonora}
that attempting to define a noncommutative gauge theory corresponding to a
subgroup of $U(n)$ is not trivially accomplished. This is
true in the case of a string-brane theory configuration. We must find a
$NCSO(3,1)$, which reduces to $CSO(3,1)$ and, ultimately, to $SO(3,1)$
when $\theta\rightarrow 0$. Bonora et al., were able to show that it is
possible to impose constraints on gauge potentials and the corresponding
gauge transformations, so that ordinary commutative orthogonal and
symplectic gauge groups were recovered when the deformation parameter
$\theta$ vanishes. These constraints were defined in terms of a generalized
gauge theory charge conjugation operator, and a generalization of
connection-based Lie algebras in terms of an antiautomorphism in the
corresponding $C^*$-algebra. The problem of deriving an
explicit description of $NCSO(d,1)$ will be investigated in a future
article.

To avoid possible problems with negative energy ghost states, we can
develop the same formalism based on the hyperbolic complex vierbein
$E^a_\mu={\rm Re}(E^a_\mu)+\omega {\rm Im}(E^a_\mu)$ and a hyperbolic
complex spin connection and curvature. The noncommutative spacetime is
defined by
\begin{equation}
\{{\hat x}^\mu, {\hat x}^\nu\}=2x^\mu x^\nu+\omega\tau^{\mu\nu},
\end{equation}
and a hyperbolic complex Moyal $\diamond$-product in terms of the pure
imaginary element $\omega$:
\begin{equation}
f(x)\diamond g(x)
=\exp\biggl[\frac{\omega}{2}\tau^{\mu\nu}\frac{\partial}{\partial\alpha^\mu}
\frac{\partial}{\partial\beta^\nu}\biggr]f(x+\alpha)g(x+\beta)\vert_{\alpha=\beta=0}
$$
$$
=f(x)g(x)+\frac{\omega}{2}\tau^{\mu\nu}\partial_\mu f(x)\partial_\nu
g(x)+{\cal O}(\tau^2).
\end{equation}

A hyperbolic complex vierbein formalism can be developed along the same
lines as the complex vierbein formalism and the invariance group of
transformations $NCSO(3,1,C)$ is extended to $NCSO(3,1,\Omega)$.

The noncommutative field equations and the action contain an infinite
number of derivatives, and so constitute a nonlocal theory at the quantum
level. When the coordinates assume their classical commutative structure,
then the theory regains standard, classical causal properties.
The nonlocal nature of the noncommutative quantum gauge field theory and
quantum gravity theory presents potentially serious problems. Such theories
can lead to instabilities, which render them
unphysical~\cite{Woodard}, and in the case of space-time
noncommutativity, $\theta^{i0}\not=0$, the perturbative S-matrix may not be
unitary~\cite{Gomis}. Moreover, it is unlikely that one can apply the
conventional, canonical Hamiltonian formalism to quantize such gauge field
theories. Of course, we know that the latter formalism already meets
serious obstacles in applications to quantum gravity for commutative
spacetime. In ref. [11], it was shown that open string theory in a 
background electric field displays space-time noncommutativity, but it was
argued that stringy effects conspire to cancel the acausal effects that are
present for the noncommutative field theory.
\section{Conclusions}

The deformed product of functions given by the
Moyal $\star$-product of functions leads to complex gauge fields and a
complex Riemannian or non-Riemannian geometry. We first considered the
possibility that the fundamental tensor $g_{\mu\nu}$ is Hermitian
nonsymmetric, leading to a nonsymmetric gravitational theory (NGT). This
theory does not at present form a viable gravitational theory, because of
instability difficulties that arise when an expansion is performed about a
GR background. The $g_{\mu\nu}$ had to be chosen to be hyperbolic
complex to avoid basic negative energy ghost problems. Even though this
hyperbolic complex nonsymmetric theory could be shown to be self-consistent
as far as asymptotic boundary conditions are concerned and to be free of
higher-order ghost poles, the instability problems discovered by
Clayton~\cite{Clayton,Clayton2} remain. It is possible that these
instability problems can be removed by some new formulation of
NGT, but so far no one has succeeded in discovering such a
formulation.

We then investigated a complex symmetric (non-Hermitian) $g_{\mu\nu}$
on a complex manifold with the local gauge group of complex
Lorentz transformations $CSO(3,1)$ (or $CSO(d,1)$ in
(d+1)-dimensions). From a real action, we obtained a consistent set of
field equations and Bianchi identities in a torsion-free spacetime. By
assuming that the physical line element was given by the real form
(\ref{realmetric}), we derived a static spherically symmetric solution
with two gravitational ``charges'' $2m$ and $2\epsilon$. When $\epsilon=0$,
we regained the standard Schwarzschild solution of GR, so there is a
well-defined classical gravity limit of the theory, which agrees with all
the know gravitational experimental data. We also considered a complex
geometry based on a hyperbolic complex metric and connection. This geometry
would avoid potential problems of negative ghost states in the linearized
equations of gravity.

By formulating the vierbein and spin connection
formalism on a flat tangent space by means of complex vierbeins and a
complex spin connection and curvature tensor, we generalized the complex
symmetric geometry to anticommuting coordinates by replacing the usual
products by Moyal $\diamond$-products and complex gauge transformations, and
by Moyal $\diamond$-products for a hyperbolic complex noncommutative
geometry. We extended the complex group of gauge transformations $CSO(3,1)$
of the commutative spacetime to a noncommutative group of complex
orthogonal gauge transformations $NCSO(3,1, C)$. For the
geometry based on a noncommutative Clifford algebra $\Omega$, the
invariance group was extended to the hyperbolic complex group
$NCSO(3,1,\Omega)$.

The CSG theory we have developed can be the basis of a quantum
gravity theory, which contains classical GR as the limit of CSG when the
new gravitational charge $\epsilon\rightarrow 0$. It would be interesting
to consider possible experimental consequences of CSG for strong
gravitational fields, for gravitational wave experiments and for
cosmology.

A noncommutative quantum gravity theory as well as a
noncommutative quantum gauge field theory pose potential
difficulties, because of the nonlocal nature of the interactions
at both the nonperturbative and perturbative levels. It is not clear whether 
a physically consistent quantum gauge field theory or quantum
gravity theory, based on noncommutative coordinates, can be found
at either the perturbative or nonperturbative levels. This is an
open question that requires more intensive investigation. It
remains to be seen whether our standard physical notions about
the nature of spacetime can be profoundly changed at the quantum
level, without leading to unacceptable physical consequences.
\vskip 0.2 true in {\bf Acknowledgments}
\vskip 0.2 true in
I thank Michael Clayton and John Madore for helpful discussions. This
work was supported by the Natural Sciences and Engineering Research Council
of Canada.
\vskip 0.5 true in


\begin{thebibliography}{100}
\bibitem{Snyder} H. S. Snyder, Phys. Rev. {\bf 71}, 38 (1947); Phys. Rev.
{\bf 72}, 68 (1947).

\bibitem{Connes} A. Connes, Noncommutative Geometry, Academic Press 1994.

\bibitem{Madore} J. Madore, An Introduction to Noncommutative Differential
Geomtry and Its Physical Applications, second edition, Cambridge University
Press, 1999.
\bibitem{Connes2} A. Connes, M. R. Douglas and A. Schwarz, JHEP 9802,
003 (1998), hep-th/9711162.

\bibitem{Witten} N. Seiberg and E. Witten, JHEP 9909, 032 (1999),
hep-th/9908142.

\bibitem{Castellani} L. Castellani, hep-th/0005210

\bibitem{Krogh} Y. K. E. Cheung and M. Krogh, Nucl. Phys. {\bf 528}, 185
(1998), hep-th/9803031.

\bibitem{Chu} C. S. Chu and P. M. Ho, Nucl. Phys. {\bf B550}, 151 (1999),
hep-th/9812219.

\bibitem{Schomerus} V. Schomerus, JHEP 9906, 030 (1999),
hep-th/99033205

\bibitem{Ardalan} F. Ardalan, H. Arfaei and M. M. Sheikh-Jabbari, JHEP
9902, 016 (1999), hep-th/9810072.

\bibitem{Seiberg} N. Seiberg, L. Susskind, and N. Toumbas, JHEP 0006,
044 (2000), hep-th/0005015 v3.

\bibitem{Seiberg2} N. Seiberg, L. Susskind, and N. Toumbas,
JHEP 0006, 021 (2000), hep-th/0005040.

\bibitem{Moyal} J. E. Moyal, Proc. Camb. Phil. Soc. {\bf 45}, 99
(1949).

\bibitem{Gomis} J. Gomis and T. Mehen, hep-th/0005129.

\bibitem{Varilly} J. C. Varilly and J. M. Gracia-Bondia, Int. J. Mod.
Phys. {\bf A14}, 1305 (1999), hep-th/9804001.

\bibitem{Martin} C. P. Martin and D. Sanchez-Ruiz, Phys. Rev. Lett {\bf
83}, 476 (1999), hep-th/9903077.

\bibitem{Wess} J. Madore, S. Schraml, P. Schupp and J. Wess,
Eur. Phys. J. {\bf C16}, 161 (2000), hep-th/0001203.
 
\bibitem{Weyl} H. Weyl, The Theory of Groups and Quantum Mechanics, Dover,
New-York (1931).

\bibitem{Chamseddine} A. H. Chamseddine, hep-th/0005222.

\bibitem{Majid} S. Majid, math.QA/0006152

\bibitem{Moffat} J. Moffat, Proc. Camb. Phil. Soc. {\bf 52}, 623 (1956).

\bibitem{Moffat2} J. Moffat, Proc. Camb. Phil. Soc. {\bf 53}, 473 (1957).

\bibitem{Moffat3} J. Moffat, Proc. Camb. Phil. Soc. {\bf 53}, 489 (1957).

\bibitem{Einstein} A. Einstein, The Meaning of Relativity, fifth edition,
Princeton University Press, 1956.

\bibitem{Straus} A. Einstein and E. G. Straus, Ann. Math. {\bf 47},
731 (1946).

\bibitem{Moffat4} J. W. Moffat, Phys. Rev. {\bf D19}, 355 (1979).

\bibitem{Moffat5} J. W. Moffat, J. Math, Phys. {\bf 21}, 1978 (1980).

\bibitem{Moffat6} J. W. Moffat, J. Math. Phys. {\bf 29}, 1655 (1988).

\bibitem{Moffat7} J. W. Moffat, Review of Nonsymmetric Gravitational
Theory, Proceedings of the Summer Institute on Gravitation, Banff Centre,
Banff, Canada, edited by R. B. Mann and P. Wesson, World Scientific,
Singapore, 1991.

\bibitem{Moffat8} J. W. Moffat, Phys. Lett. {\bf B355}, 447 (1995),
gr-qc/9411006.

\bibitem{Moffat9} J. W. Moffat, J. Math. Phys. {\bf 36}, 3722 (1995).

\bibitem{Damour} T. Damour, S. Deser and J. McCarthy, Phys. Rev. {\bf D47},
1541 (1993).

\bibitem{Clayton} M. A. Clayton, J. Math. Phys. {\bf 37}, 395 (1996).

\bibitem{Clayton2} M. A. Clayton, Int. J. Mod. Phys. {\bf A12}, 2437
(1997).

\bibitem{Clifford} W. K. Clifford, Proc. London Math. Soc. {\bf 4}, 381
(1873); Am. J. Math. {\bf 1}, 350 (1878); I. M. Yaglom, Complex numbers in
Geometry, Academic Press, London, 1968.

\bibitem{Bonora} L. Bonora, M. Schnabel, M. M. Sheik-Jabbari, and A.
Tomasiello, hep-th/0006091.

\bibitem{Woodard} D. A. Eliezer and R. P. Woodard, Nucl. Phys. {\bf B325},
389 (1989); R. P. Woodard, hep-th/0006207; J. Gomis, K. Kamimura and J. 
Llosa, hep-th/0006235.

\end{thebibliography}
\end{document}